\documentclass[prl,floats,aps,showpacs,twocolumn]{revtex4-1}
\usepackage{graphicx}
\usepackage{dcolumn}
\usepackage{bm}

\begin{document}

\title{Gravitational Wave Emission from the Single-Degenerate Channel of Type Ia Supernovae}
\author{David Falta$^1$}
\author{Robert Fisher$^1$}
\email{robert.fisher@umassd.edu}
\author{Gaurav Khanna$^{1,2}$}
\affiliation{$^1$Physics Department, University of Massachusetts Dartmouth, \\ 285 Old Westport Road, North Dartmouth, MA 02747-2300}
\affiliation{$^2$Max-Planck-Institut f\"{u}r Gravitationsphysik, Hannover, Germany}
\date{\today}

\begin{abstract}
The thermonuclear explosion of a C/O white dwarf as a Type Ia
supernova (SN Ia) generates a kinetic energy comparable to that
released by a massive star during a SN II event. Current observations
and theoretical models have established that SNe Ia are asymmetric,
and therefore--like SNe II--potential sources of gravitational wave
(GW) radiation.  We perform the first detailed calculations of the GW
emission for a SN Ia of any type within the single-degenerate channel.
The  gravitationally-confined detonation (GCD) mechanism predicts a
strongly-polarized GW burst in the frequency band around 1 Hz.
Third-generation spaceborne GW observatories currently in planning may
be able to detect this predicted signal from SNe Ia at distances up to
1 Mpc. If observable, GWs may offer a direct probe into the first few
seconds of the SNe Ia detonation.
\end{abstract}



\maketitle

\textit{Introduction.}---The discovery of the Phillips relation \citep{phillips93} enabled the use of Type Ia supernova (SN Ia) as standardizable cosmological candles, and ushered in a new era of astronomy leading to the discovery of the acceleration of the universe \citep{riessetal98, perlmutteretal99}. Understanding the evolution of SNe Ia may improve the systematics errors associated with the the calibration of the Phillips relation and allow precision studies of dark energy \citep{hoflichetal10}.

The single-degenerate (SD) channel of SNe Ia consists of a C/O white dwarf (WD) accreting mass from a non-degenerate main sequence or red giant companion. As the WD approaches the Chandrasekhar limit, carbon burning is initiated in its convective core \citep{whelanetal73,nomotoetal84}. After a few hundred years of this ``simmering'' phase, unstable thermonuclear burning is expected to ignite at  one or more off-centered points, giving rise to a buoyantly-rising, subsonic deflagration flame bubble.

\begin{figure}[H]
\includegraphics[scale=1.0]{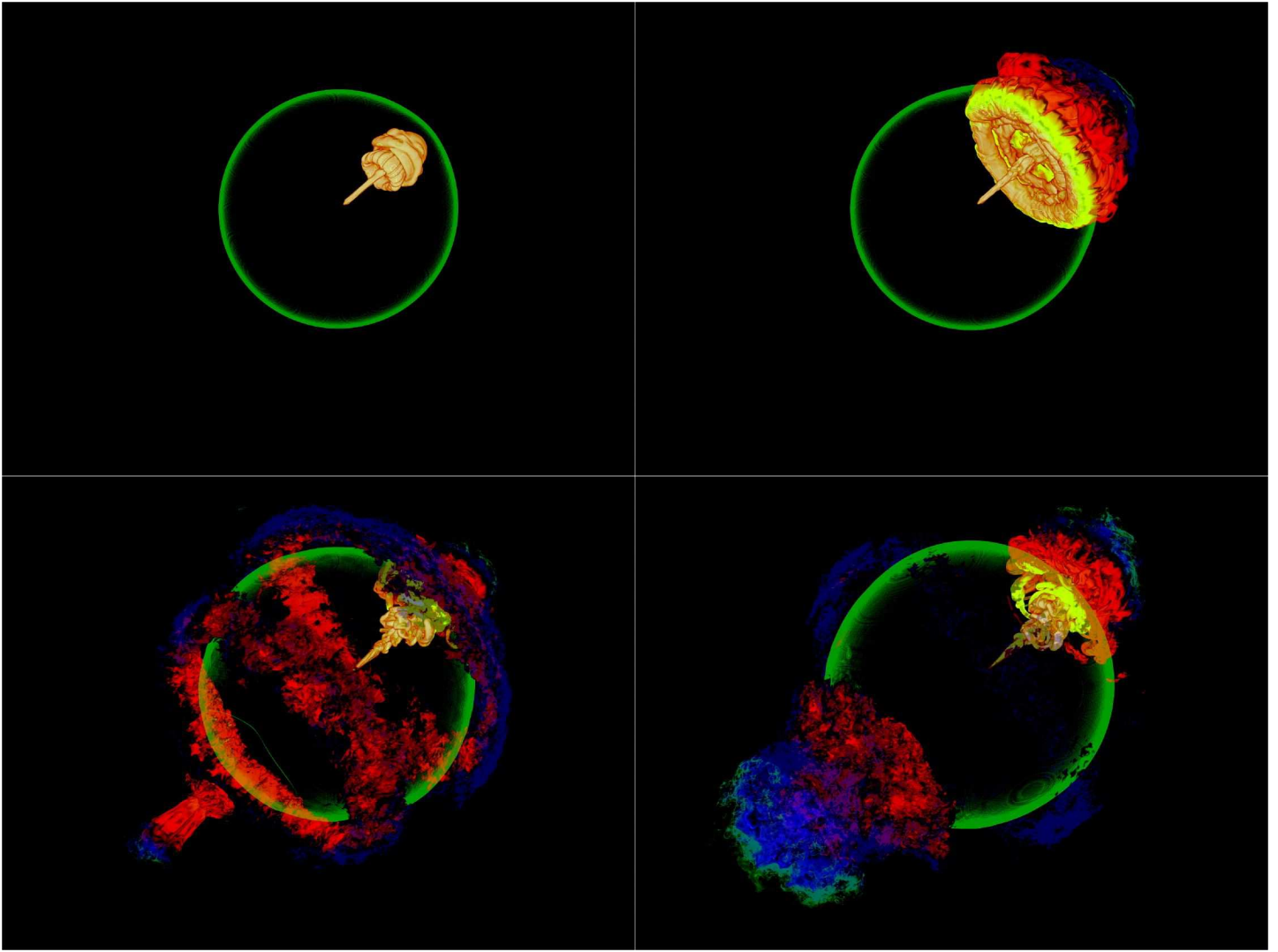}
\caption{Time evolution of the early phases of the development of the GCD from breakout through initiation of detonation. The ash temperature field is volume rendered, and the surface of the white dwarf is shown in green.The top left-frame shows the flame bubble just prior to breakout at $t = 0.9$ s. The top right shows the bubble just after breakout, depicting the ash surface flow over the gravitationally-bound WD. The lower-left frame shows the development of the jet at the point opposite to breakout. The final frame shows the initiation of the jet at $\sim 2.5$ s. Figure reproduced from \cite {jordanetal08} by permission of the AAS.}
\label{defdet}
\end{figure}

Both the deflagration-to-detonation transition (DDT) \cite{khokhlov91} and the gravitationally-confined detonation (GCD) \cite{plewaetal04, jordanetal08} models for the SN Ia explosion mechanism begin with this stage of off-centered ignition. The predictions of the two mechanisms depart by the evolution of the bubble as it approaches breakout. The DDT mechanism posits that a transition from deflagration to detonation is made prior to bubble breakout. While the DDT mechanism yields results consistent with observation, it requires that the density at which the transition is made to be set as a free parameter. In contrast, in the GCD model, the flame bubble breaks through the surface of the WD, launching ash into a surface flow; see Figure \ref {defdet}. The initial deflagration phase is relatively inefficient, typically burning only a modest fraction of the mass of the WD. Therefore, in the GCD model, the WD remains gravitationally-bound, leading to a ram-pressure driven detonation at the point opposite of bubble breakout, which subsequently unbinds the WD. Significantly, because both the GCD and the DDT initiate detonations near the edge of the WD, both are intrinsically and strongly asymmetrical. 

Spectropolarimetry measurements of lines of intermediate mass ions, including Si II and Ca II, yield greater polarizations in the outer layers of SNe Ia \citep {wangwheeler08}, strongly suggesting asymmetry in the later phases of burning. Furthermore, \citet{maedaetal10} demonstrated that an intrinsic asymmetry in Ia ejecta, viewed from a random direction, can account for their spectral evolution diversity. Taken together, these observations are broadly consistent with the asymmetries predicted by the DDT and the GCD mechanisms of SNe Ia. The combined intrinsic asymmetry and large kinetic energy of the explosion motivates us to consider the nature of gravitational wave (GW) radiation from SNe Ia in this paper. As is the case with SNe II, which are known to be candidate GW sources in the $10^2 - 10^3$ Hz region of LIGO-like instruments \cite {niemeyeretal96, muelleretal97, ott09}, the asymmetry of SNe Ia naturally leads to the production of GWs. 

We may simply estimate the GW signal strength of a SN Ia using the Newtonian-quadrupole approximation to the Einstein field equations: $h \simeq G \ddot{Q}/ (c^4 D) $. Here $Q$ is the quadrupole mass moment, $D$ is the distance to the source, and $G$ and $c$ are the universal gravitational constant, and the speed of light, respectively. \textit{We can place a robust upper-bound on the SN Ia GW amplitude by assuming that the Ia mechanism is highly aspherical, such that the non-spherical kinetic energy is} $ \ddot{Q} / 4\sim E_{\rm kin}^{\rm ns} \sim E_{\rm kin} \sim 10^{51}$ ergs \cite {thorne98}. With this assumption, we estimate an upper-bound of the dimensionless strain amplitude as $10^{-20}$ at a distance of 10 kpc.

We next estimate the characteristic GW frequency range of a SNe Ia detonation. The detonation speed $v_{\rm det}$ within a WD of radius $R$ is set according to the Chapman-Jouguet detonation condition \citep{fickettanddavis79}, and yields a characteristic frequency $f_{\rm det} \sim v_{\rm det} / (2 R)$. Estimating the radius $R \sim 2000$ km as that of a cold, near-Chandrasekhar mass C/O WD, and the Chapman-Jouguet speed as $v_{\rm det} \sim 10^4$ km/s, we establish an upper-bound to the frequency of the expected GW signal at $\lesssim 2.5$ Hz. This simple estimate implicitly assumes no pre-expansion occurs during the initial deflagration phase, and corresponds to an overluminous SNe Ia event which will yield $\simeq 1.2 M_{\odot}$ of $^{56}$Ni \citep {meakinetal09}.   A lower-bound to the detonation frequency can be estimated using the pre-expansion out to $R \sim 4000$ km required to produce the observed nucleosynthetic yield of  $\simeq 0.7 M_{\odot}$ $^{56}$Ni in normal brightness Ia events \cite {mazzalietal07} : $f_{\rm det} \gtrsim 1$ Hz.

If the detonation is sufficiently asymmetric, this GW signal may be detectable by proposed third-generation  GW instruments. We next develop a more refined estimate by post-processing three-dimensional (3D) simulations of the GCD mechanism \cite {jordanetal08}.

\textit{Methodology.}---We simulate the evolution of a 3D hydrodynamical GCD SNe Ia in the SD channel \citep{jordanetal08}. The progenitor WD is evolved from the onset of deflagration through detonation. The simulations employ the Euler equations of inviscid, non-relativistic hydrodynamics, coupled to Poisson's equation for self-gravity, and an advection-diffusion reaction model of the thickened combustion front. The non-relativistic dynamical assumption is well-justified here, because the highest bulk fluid speeds achieved during the simulation are of order the Chapman-Jouguet speed of the supersonic detonation front, about $\sim 0.03c$, and the gravitational redshift factor $\sim G M / (c^2 r) \sim 1 \cdot 10^{-6}$ at the surface of the WD.

The simulations utilize FLASH -- a modular, component-based application code framework created to simulate compressible, reactive astrophysical flows \citep{fryxelletal00}. The framework supports a block-structured adaptive mesh refinement (AMR) grid, and includes a directionally-split inviscid hydrodynamic piecewise-parabolic method (PPM) solver \citep{colellawoodward84}. A Helmholtz equation of state describes the thermodynamic properties of the stellar plasma, including contributions from blackbody radiation, ions, and electrons of an arbitrary degree of quantum degeneracy \cite{timmes2000a}. A multipole solver method determines the gravitational potential from Poisson's equation. Three scalar progress variables track the flames and detonation waves, following the stages of carbon burning, quasi-static equilibrium (QSE) relaxation, and nuclear statistical equilibrium (NSE) relaxation \citep{calder2007,townsley2007}. 

The simulation begins with a C/O WD, with a single flame bubble ignited slightly off-center.
Previous work on the GCD has demonstrated robust detonations, arising independent of the resolution in the detonation region, and for a wide variety of initial bubble sizes and offsets \citep{jordanetal08, meakinetal09, seitenzahletal09}. The simulation we present begins with a 1.37 $M_\odot$ C/O WD, with a single pre-ignited flame bubble of radius 16 km, and 40 km offset from the center of the WD. We use a $(49152 \mathrm{km})^3$ computational mesh that ensures boundary effects have no influence during detonation expansion. To ensure convergence in the GW signal we have performed runs at maximum finest resolutions of  8km, 12km, and 16km. 

We employ the standard post-Newtonian expressions for the gravitational quadrupole radiation field from a 3D source. In particular, we calculate the dimensionless GW amplitudes $h_+$ and $h_\times$, which correspond to the two independent polarizations in the transverse-traceless (TT) gauge \citep{nakamuraetal89}. We utilize an expression for the second-time derivative of the reduced quadrupole mass moments $I_{ij}$, in which the time-derivatives are reduced to spatial derivatives using the mass continuity equation, and the hydrodynamical Euler equations \citep{nakamuraetal89}. This method is formally one order higher accuracy than evaluating the time derivative of the quadrupole moment twice, as shown in \cite{moenchmeyeretal91}.

\begin{figure}[H]
\includegraphics[scale=0.75]{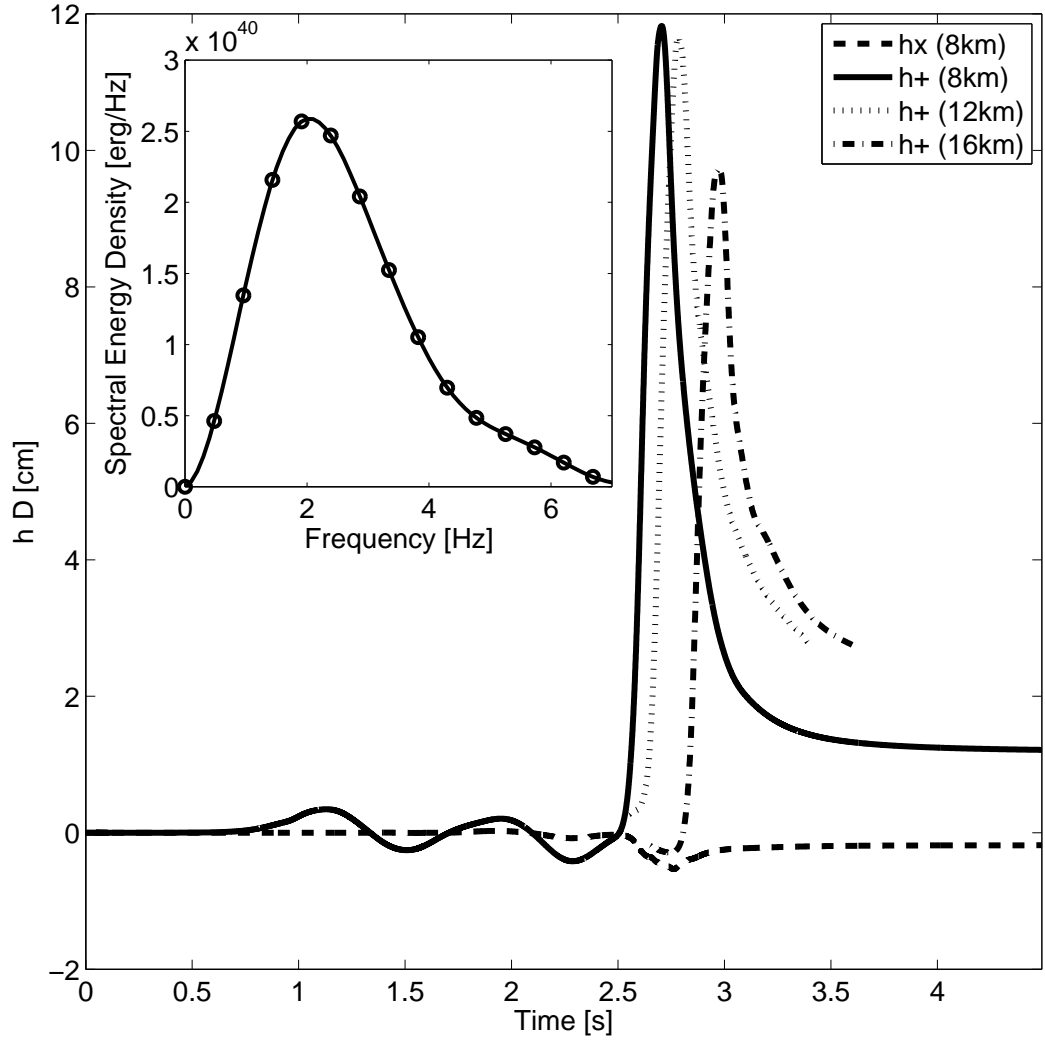}
\caption{$h_+$ and $h_\times$ amplitudes when viewed on the equator ($\theta=\frac{\pi}{2}$, $\phi=0$). Inset: Spectral energy density. }
\label{signal}
\end{figure}

\textit{Results.}---We present in Figure \ref{signal} the product of the dimensionless GW signal strength $h$ and the distance to the source $D$, for observers with line of sight on the equator ($\theta=\pi / 2, \phi=0$). The simulation begins at $t = 0$ with the flame bubble ignition. The first $\sim$ 2.5 s after bubble ignition correspond to the deflagration phase; see Figure \ref {defdet}. During this phase, the flame bubble rises and breaks out of the surface of the WD at $\sim$0.9 s leaving behind a ``stem'' that continues to burn beyond breakout. The deflagration phase continues as the ash moves over the surface of the WD, and powers a jet that gives rise to a detonation front which is initiated at a point opposite of bubble breakout at $\sim$ 2.5 s. 

During detonation at $\sim$2.5-3 s, the supersonic detonation front  unbinds the WD, and releases a kinetic energy $E_{\rm kin}\sim 1.4 \cdot 10^{51}$ ergs and a nuclear energy $E_{\rm nuc}\sim 1.34 \cdot 10^{51}$ ergs. The GW signal is dominated by the $h_+$ polarization, while the $h_\times$ polarization deviates only slightly from zero due to the growth of non-axisymmetric instabilities. The $h_+ D$ signal is largely shaped by simple geometric considerations; it is nearly symmetric in time about 2.7 s,  when the signal reaches a maximum as the detonation front passes over the center of the WD. The $h_+ D$ signal achieves a maximum amplitude of 11.8 cm before undergoing two phases of decline. During the first phase, from  $\sim$2.7-3 s, the detonation front propagates over the second hemisphere of the WD. During the second phase, at $t > $3 s, the detonation has completely compressed the WD, and the ejecta freely expand into the homologous phase of the explosion. 

\begin{figure}[H]
\includegraphics[scale=1.0]{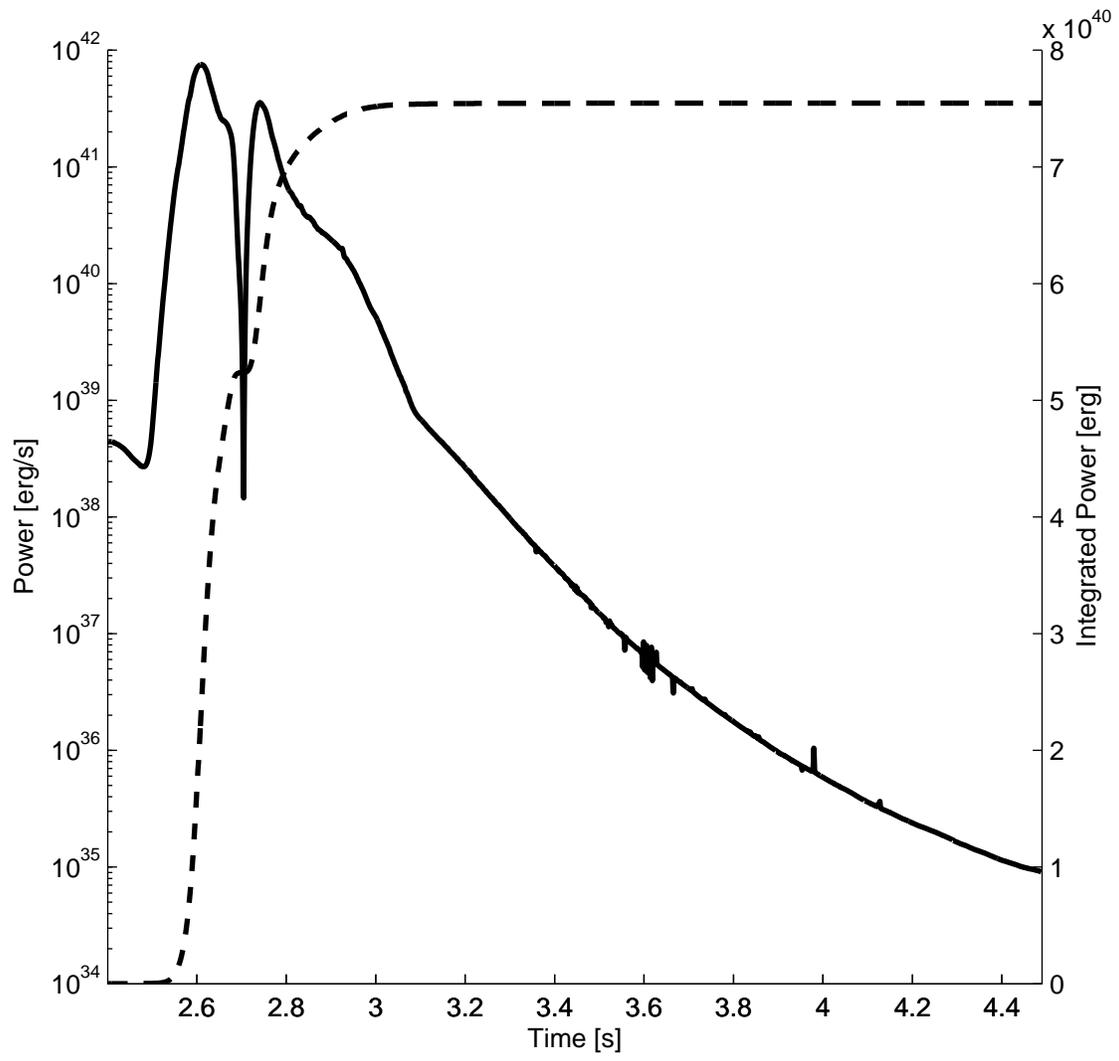}
\caption{Power radiated (solid) and integrated power (dashed) of emitted GWs.}
\label{power}
\end{figure}

An analysis of the GW spectral energy density over all angles, figure \ref{signal} (inset), shows a characteristic frequency near 2 Hz, which is consistent with our estimate. Integration yields the total energy released as GW radiation to be $7.5 \times 10^{40}$ ergs ($4.2 \times 10^{-14}$ of $M_{\odot}c^2$). A time analysis of the radiated GW power and the cumulative radiated energy, figure \ref{power}, establishes that all appreciable GW energy is released during detonation, by $\sim 3$ s. 

We conducted a convergence study, keeping the initial conditions and parameters of the runs fixed, and adding higher levels of maximum refinement, corresponding to finest resolutions of 16 km, 12 km, and 8 km (figure \ref{signal}). In terms of the peak  $h_+ D$ signal amplitude at the equator, relative to our highest-resolution 8 km run, our 12 km and 16 km runs are converged to within within 1.1\% and 19.1\%, respectively. 

\begin{figure}[H]
\includegraphics[scale=1.0]{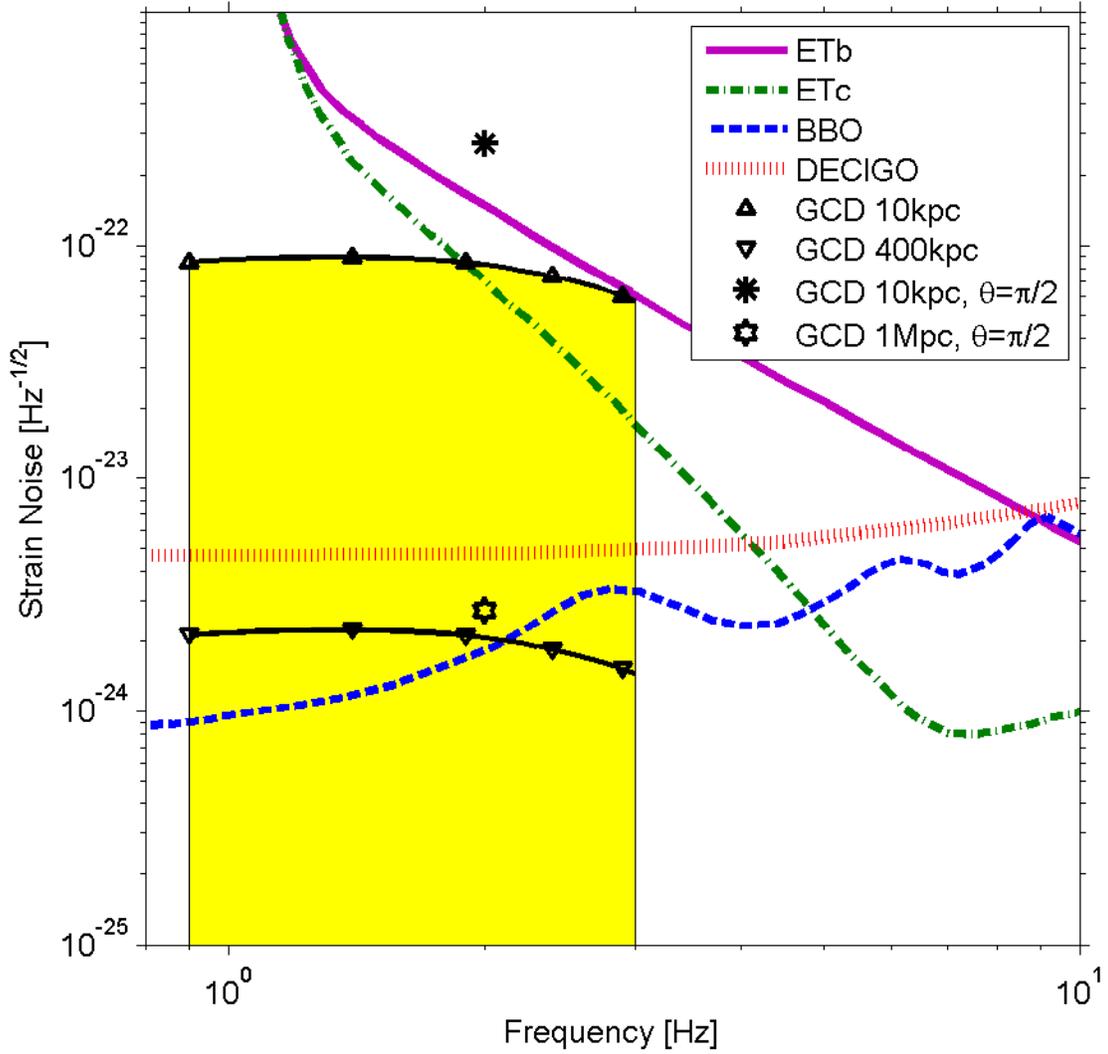}
\caption{Strain noise curves for current proposed detectors sensitive to GCD sources. The filled region represents possible GCD event signals in the range between 0.9 Hz and 3.0 Hz, corresponding to significant pre-expansion and overluminous events respectively. The region is capped by the characteristic signal strength at 10 kpc for a randomly-oriented source. We also plot the maximum GCD amplitudes at 10 kpc and 1 Mpc, which occur when viewed along the equator ($\theta = \pi / 2$).}
\label{strain_detect}
\end{figure}

In figure \ref{strain_detect}, we compare the characteristic GW strain noise for our calculated GCD SNe Ia events with some third-generation GRW detectors currently in planning,  including the Big Bang Observer (BBO) \citep{phinneyetal04}, Deci-Hertz Interferometer Gravitational Wave Observatory (DECIGO) \citep{kawamuraetal06}, or the Einstein Telescope (ET) a and b\citep{hildetal10}. Detection of nearby extragalactic events out to 1 Mpc could be possible by BBO if the event happens to be oriented such that line of sight is close to the equator. A more conservative estimate, based upon randomly-oriented events, suggests that extragalactic events would be observable by BBO out to about 400 kpc.

\textit{Discussion.}--- Our single-bubble GCD models produce intermediate mass elements at a velocity coordinate $\sim 11,000$ km/s, creating a layered structure of IME and Fe peak (NSE) products similar to observation \cite{mazzalietal07, meakinetal09}. However, single-bubble GCD models generally underproduce intermediate mass elements and overproduce $^{56}$Ni. Consequently, the models are generally too luminous in comparison to Branch normal Ia events. The Ni-overproduction issue may be resolved by beginning with multiple simultaneous ignition points \cite{jordanetal09}, or through alternative subgrid models of turbulent nuclear burning, which will effectively enhance the turbulent burning rate during the deflagration phase, and lead to pre-expansion of the WD progenitor. Because the later detonation phase establishes the kinetic energy of the explosion in the GCD model, we do not expect that enhanced pre-expansion will significantly impact the strengths of the predicted GW signals presented here. Pre-expansion will, however, slightly shift the characteristic GW frequency of the GCD mechanism downward by a factor of $\sim$ 2. 

Our current theoretical models predict signal strengths which would be detectable for galactic SNe Ia, which are relatively rare, with event rates of one per few hundred years. However, like early models of SNe II, current Ia models have yet to include rotation, convection, and other intrinsically multi-dimensional effects which can be expected to yield greater signal strengths than predicted here. Simultaneous detection in the optical of nearby extragalactic SNe within $\sim$ few Mpc with concurrent measurements, including non-detections, from future GW observatories may ultimately yield useful constraints upon these future models.

GK and DF acknowledge support from NSF grant numbers PHY-0902026 and PHY-1016906. The software used in this work was in part developed by the DOE-supported ASC / Alliance Flash Center at the University of Chicago. This research was supported in part by the National Science Foundation through TeraGrid resources provided by the  Louisiana Optical Network Initiative under grant number TG-AST100038.

\newcommand{\aj}{Astronomical Journal}
\newcommand{\apjl}{Astrophysical Journal Letters}
\newcommand{\apjs}{Astrophysical Journal Supplement}
\newcommand{\aap}{Astronomy and Astrophysics}
\newcommand{\araa}{Annual Review of Astronomy and Astrophysics}
\newcommand{\mnras}{Monthly Notices of the Royal Astronomical Society}

\end{document}